# On the similarity analysis of the growth of microalgae

Chunyang Ma*

*School of Advanced Manufacturing, Nanchang University, 999 Xuefu Avenue, Nanchang 330031,*

*People's Republic of China*

*School of Mechanical and Electrical Engineering, Nanchang University, 999 Xuefu Avenue,*

*Nanchang 330031, People's Republic of China*

**Abstract**

Similarity analysis of microalgae growth is crucial for understanding microalgae cultivation, such analysis has not been considered before in previous studies. This letter considers the natural features of the microalgae growth phenomenon that are different from the inanimate process. Accordingly, the similarity of light transfer within microalgae suspensions was comprehensively studied. The characteristic number of microalgae growth and the temporal scaling function were considered to represent the growth relate characteristics. Subsequently, a group of dimensionless numbers were derived to characterize the time-dependent similarity of light within microalgae suspensions. For growth similarity, the newly introduced characteristic number of microalgae growth and the temporal scaling function of extinction, absorption and scattering were considered identical. The similarity relations of the time-dependent optical constants of microalgae were also provided for similar growth. The similarity analysis of the microalgae growth will be useful in designing and guiding experiments on microalgae cultivation.

**Keywords:** Similarity principle, Temporal scaling function, Characteristic number, Optical constants, Microalgae

---

*Corresponding author.

*E-mail address*: cyma@ncu.edu.cn, cymaphy@163.com.

## 1. Introduction

As an alternative form of energy, microalgae can be applied to numerous fields, such as the pharmaceutical industry, environmental protection, space food, etc. [1-3]. The microalgae production process can fix carbon dioxide via cell photosynthesis. There is no net carbon dioxide emission during the cycle of microalgal biomass. Therefore, it can facilitate the critical objective of achieving carbon neutrality [4, 5].

Early studies on microalgae production dates back to 1960s [6]. At present, there is much more knowledge about microalgae photosynthesis and production. However, the cultivation costs remain high, which seriously limits the commercial application of microalgae [7, 8]. Moreover, experimental research data of microalgae is extremely diverse for different cultivation conditions and microalgae strains. All experimental measurements of microalgae were under specific cultivation conditions, such as light intensity, nutrients concentration, and temperature [9-12]. On the one hand, excessive light supply can lead to photoinhibition. On the other hand, insufficient light is detrimental to the growth of microalgae. Accordingly, the optimum light intensity is different for different microalgae strains [13, 14]. The effect of different nitrogen sources on the growth of microalgae has been experimentally verified [15]. The supply of nutrients also exhibits a complex impact the growth of microalgae [16, 17]. Numerous theoretical growth models of microalgae have been established based on the phenomena observed from experiments [18-20]. However, selecting an adaptive growth model for a specific situation is not a simple problem. Generally, previous studies on microalgae have significantly advanced the understanding and application of microalgae.

Admittedly, we have measured a considerable volume of data on microalgae. However, it is still difficult to comprehend and explain the measured data under a unified theoretical frame. The theoretical difficulty leads to negative effects on the study of microalgae. For instance, it leads to underutilizing the experimental data and not fully comprehending the representation of the data. This dilemma has been persistent, which could fundamentally be owing to the growth of microalgae cells. A similar problem was also encountered in heat transfer science, which was successfully solved after the introduction of the similar principle and dimensionless theorem [21, 22]. Inspired by this, this letter attempts to extend the applicational range of the similar principle to microalgae growth. In short, the aim is to propose a unified perspective to explain and understand the experimental data on microalgae growth.

In this letter, the unified theory frame of microalgae growth will be preliminarily established based on the similar principle. The theory will be used to the design and guide the experiment of microalgae cultivation. It will help simplify the complexity of the diverse measured data at the former perspective. The measured data can be represented as a similar group under the perspective of the unified theory. Moreover, the work will deepen the understanding of microalgae growth.

## 2. Similarity analysis of microalgae growth

Microalgae growth is a multi-physics problem, which includes the light radiation, nutritional concentration, fluid, energy, and temperature fields. In accordance with the similarity theory, physical processes are similar if their characteristic numbers are equal [23].

$$\text{Sc}'=\text{Sc}'',\ \text{Sh}'=\text{Sh}'',\ \text{Re}'=\text{Re}'',\ \text{Eu}'=\text{Eu}'',\ \text{Nu}'=\text{Nu}'',\ \text{Pr}'=\text{Pr}'' \qquad (1)$$

For concentration, fluid, and temperature fields, respectively, and the conditions for unique solution are similar, including geometric, boundary, physical, and initial conditions. For a considered problem, the characteristic numbers should be reasonably selected from the existing group of characteristic numbers. For example, the Grashof number, $\text{Gr}$, should be chosen if natural convection is considered.

We will analyze the similarity problem for the growth of microalgae., The starting point will be the radiative transfer equation (RTE) [24]

$$\frac{n}{c}\frac{\partial I(\mathbf{r},\mathbf{s})}{\partial t}+\mathbf{s}\cdot\nabla I(\mathbf{r},\mathbf{s})+\beta I(\mathbf{r},\mathbf{s})=\frac{\kappa_s}{4\pi}\int_{\bar{\Omega}=4\pi}I(\mathbf{r},\bar{\mathbf{s}})\Phi(\bar{\mathbf{s}},\mathbf{s})\mathrm{d}\bar{\Omega}+\kappa_a I_b \qquad (2)$$

where $I$ denotes the radiative intensity in the direction $\mathbf{s}$ at location $\mathbf{r}$, $\mathbf{s}$ corresponds to the direction vector, $\kappa_s$ symbolizes the scattering coefficient, $\beta=\kappa_a+\kappa_s$ indicates the extinction coefficient, $\kappa_a$ denotes the absorption coefficient, $\Phi(\bar{\mathbf{s}},\mathbf{s})$ refers to the scattering phase function, and $\bar{\Omega}$ is the solid angle. Note that the RTE is a function of wavelength. The similarity analysis on the microalgae

growth will be based on the RTE. A a series of dimensionless parameters is introduced as follows:

$$\frac{x'}{x''} = \frac{y'}{y''} = \frac{z'}{z''} = \frac{l'}{l''} = C_l, \quad \frac{t'}{t''} = C_t, \quad \frac{\kappa_a'}{\kappa_a''} = C_{\kappa_a}, \quad \frac{\kappa_s'}{\kappa_s''} = C_{\kappa_s},$$

$$\frac{\beta'}{\beta''} = C_\beta, \quad \frac{\Phi'}{\Phi''} = C_\Phi, \quad \frac{I'}{I''} = C_I, \quad \frac{I_b'}{I_b''} = C_b, \quad \frac{n'}{n''} = C_n \quad (3)$$

where the superscript denotes different microalgae growth. The RTE equation in the $'$ frame can be written as

$$\frac{n'}{c}\frac{\partial I'(\mathbf{r}',\mathbf{s}')}{\partial t'} + \mathbf{s}' \cdot \nabla' I'(\mathbf{r}',\mathbf{s}') + \beta' I'(\mathbf{r}',\mathbf{s}') = \frac{\kappa_s'}{4\pi}\int_{4\pi} I'(\mathbf{r}',\overline{\mathbf{s}})\Phi'(\overline{\mathbf{s}},\mathbf{s}')d\overline{\Omega} + \kappa_a' I_b' \quad (4)$$

Substituting Eq.(3) into the Eq.(4), yields the following expression:

$$\frac{C_n C_I}{C_t}\frac{n''}{c}\frac{\partial I''}{\partial t''} + \frac{C_I}{C_l}\mathbf{s}'' \cdot \nabla'' I'' + C_\beta C_I \beta'' I'' = C_{\kappa_s} C_I C_\Phi \frac{\kappa_s''}{4\pi}\int_{4\pi} I''\Phi'' d\overline{\Omega} + C_{\kappa_a} C_b \kappa_a'' I_b'' \quad (5)$$

where the dumb indicator $\overline{\Omega}$ denotes the invariance under the transformation, $\mathbf{s}'$ equals to $\mathbf{s}''$, and the parameters in parentheses are omitted here for brevity. When comparing Eq.(4) and Eq.(5), the similarity theory requires the following relationship to hold

$$\frac{C_n C_I}{C_t} = 1, \quad \frac{C_I}{C_l} = 1, \quad C_\beta C_I = 1, \quad C_{\kappa_s} C_I C_\Phi = 1, \quad C_{\kappa_a} C_b = 1 \quad (6)$$

Substituting Eq.(3) into Eq.(6), yields

$$\frac{n'}{n''}\frac{I'}{I''}\left(\frac{t'}{t''}\right)^{-1} = 1, \quad \frac{I'}{I''}\left(\frac{l'}{l''}\right)^{-1} = 1, \quad \frac{\beta'}{\beta''}\frac{I'}{I''} = 1, \quad \frac{\kappa_s'}{\kappa_s''}\frac{I'}{I''}\frac{\Phi'}{\Phi''} = 1, \quad \frac{\kappa_a'}{\kappa_a''}\frac{I_b'}{I_b''} = 1 \quad (7)$$

Following a series of algebraic calculations, the following equations can be obtained:

$$\frac{n'}{\beta' t'} = \frac{n''}{\beta'' t''}, \quad \beta' l' = \beta'' l'', \quad \beta' I' = \beta'' I'', \quad \kappa_s' l' \Phi' = \kappa_s'' l'' \Phi'', \quad \kappa_a' I_b' = \kappa_a'' I_b'' \quad (8)$$

The similarity of microalgae growth indicates that the above equations should be satisfied. Currently, we obtained a set of similarity relations. However, we need to discuss in depth what this set of relationships mean. For the convenience of discussion, The RTE equation is made dimensionless by choosing the characteristic length, $L$, characteristic time, $t_c$, and the characteristic radiative intensity, $I_{inc}$, respectively, we obtain

$$\frac{n}{c}\frac{L}{t_c}\frac{\partial I^*(\mathbf{r},\mathbf{s})}{\partial t^*}+\mathbf{s}\cdot\nabla^* I^*(\mathbf{r},\mathbf{s})+\beta L I^*(\mathbf{r},\mathbf{s})=\frac{\kappa_s L}{4\pi}\int_{\bar{\Omega}=4\pi} I^*(\mathbf{r},\bar{\mathbf{s}})\Phi(\bar{\mathbf{s}},\mathbf{s})\mathrm{d}\bar{\Omega}+\kappa_a L I_b^* \qquad (9)$$

where $\tau=\beta L$ denotes the dimensionless optical thickness, superscript $*$ stands for the dimensionless quantity. The terms $\kappa_s L$, $\kappa_a L$ can be defined as scattering thickness, $\tau_s$ and absorption thickness, $\tau_a$, respectively, and $\tau=\tau_s+\tau_a$. Introducing the term $\frac{n}{c}\frac{L}{t}\equiv\varphi$ defines dimensionless time. The similarity relations in Eq.(8) can be rewritten as follows:

$$\varphi'=\varphi'',\quad \tau'=\tau'',\quad \frac{1}{1-\omega'}f'=\frac{1}{1-\omega''}f'',\quad \omega'\Phi'=\omega''\Phi'',\quad \tau'_\eta=\tau''_\eta \qquad (10)$$

where $\omega$ symbolizes the scattering albedo, the dimensionless function $f$ is defined as $f(I,I_b)=I/I_b$, which represents the dimensionless radiation intensity with respect to the intensity of the emit term. The last similarity relation of Eq.(10) chooses the $\eta I_b(T)$ as a reference to dimensionless, where $\eta$ denotes the wavenumber, which is also regarded as a characteristic length, $I_b(T)$ satisfies the renowned *Planck's law* (differ by a constant factor of $\pi$, $I_b(T)=E_{b\lambda}/\pi$), and $T$ indicates the Kelvin temperature.

Dimensionless similarity of Eq.(10) means that the dimensionless time, optical thickness, ratio of dimensionless intensity to 1 minus scattering albedo, product of the scattering albedo multiplication the scattering phase function, and optical thickness of characteristic wavelength should be equal at the condition of similarity growth of microalgae. However, is it necessary to fully consider these five pair relations in similarity problem of microalgae growth? The microalgae growth process can be regarded as steady state compared to the light transport process, which is due to the high velocity of light. Moreover, the microalgae cells primarily absorb light in the visible spectrum. Therefore, microalgae cultivation does not need to consider the transient and emit terms in the RTE.

Considering the growth dependent radiative properties of microalgae, the equations $\tau' = \tau''$, $\omega'\Phi' = \omega''\Phi''$ can be rewritten as follows:

$$\frac{C'_{ext}N'}{C''_{ext}N''} = \frac{l''}{l'}, \quad \frac{C'_{sca}N'\Phi'}{C''_{sca}N''\Phi''} = \frac{l''}{l'} \tag{11}$$

the growth radiative properties are a function of time, which is omitted for brevity. The growth dependent radiative properties of microalgae satisfy the temporal scaling law [25, 26]

$$\frac{C_{ext}(t)}{C_{ext}(t_{sta})} = Z_e, \quad \frac{C_{sca}(t)}{C_{sca}(t_{sta})} = Z_s \tag{12}$$

where $Z_e$, and $Z_s$ represent the temporal scaling function of extinction and scattering, respectively. Substituting Eq.(12) into Eq.(11) yields

$$\frac{C'_{ext}(t_{sta})Z'_e N'}{C''_{ext}(t_{sta})Z''_e N''} = \frac{l''}{l'}, \quad \frac{C'_{sca}(t_{sta})Z'_s N'\Phi'}{C''_{sca}(t_{sta})Z''_s N''\Phi''} = \frac{l''}{l'} \tag{13}$$

Eq.(13) can be rewritten as follows:

$$\frac{Z'_e}{Z''_e} = \frac{C''_{ext}(t_{sta})N''l''}{C'_{ext}(t_{sta})N'l'}, \quad \frac{Z'_s}{Z''_s} = \frac{C''_{sca}(t_{sta})N''\Phi''l''}{C'_{sca}(t_{sta})N'\Phi'l'} \tag{14}$$

Using the extinction and scattering coefficients, Eq.(14) can be rewritten as follows:

$$\frac{Z'_e}{Z''_e} = \frac{\beta''(t_{sta})l''}{\beta'(t_{sta})l'}, \quad \frac{Z'_s}{Z''_s} = \frac{\kappa''_s(t_{sta})\Phi''l''}{\kappa'_s(t_{sta})\Phi'l'} \tag{15}$$

The similarity of microalgae growth necessitates the growth dependent radiative properties to satisfy the similarity principle

$$Z'_e = Z''_e, \quad Z'_s = Z''_s \tag{16}$$

In other words, Eq. (16) indicates that temporal scaling function of extinction and scattering is identical, for microalgae growth obey the similarity principle. And vice versa, the similarity relations of Eq.(10) can also derived from the similarity of the temporal scaling function. The same conclusion is drawn for the term, $f'(1-\omega')^{-1} = f''(1-\omega'')^{-1}$.

Let us expand the analysis of the emission term, for example, for algal cells that can absorb infrared light

$$C'_{abs}N'I'_b = C''_{abs}N''I''_b \tag{17}$$

Combining the temporal scaling law, we obtain

$$C'_{abs,sta}Z'_aN'I'_b = C''_{abs,sta}Z''_aN''I''_b \tag{18}$$

Rewritten as

$$\frac{Z'_a}{Z''_a} = \frac{C''_{abs,sta}N''I''_b}{C'_{abs,sta}N'I'_b} \tag{19}$$

The similarity of microalgae growth requires the following relation to hold

$$Z'_a = Z''_a \tag{20}$$

The relation of the emit term combined with *Planck's law* under the condition for microalgae suspension at thermal equilibrium can be expressed as follows:

$$\frac{\kappa'_a}{\kappa''_a} = \frac{T''n''^3}{T'n'^3} \tag{21}$$

where the condition $\frac{hc_0}{n\lambda kT} \ll 1$ is used, or Planck's constant $h \to 0$, i.e., postulating by classical statistics, photons of arbitrarily small energy content can be emitted. The temperature is inversely proportional to the absorption coefficient, this means that the absorption coefficient will be decreased by increasing the cultivation temperature under the condition of similarity of microalgae growth.

Until now, the growth of microalgae has not been directly considered. The similarity of the external influence field conditions on the microalgae growth are fully considered in the above analysis. The characteristic number describing the microalgae growth, which is called Microalgae growth number, $Mg$, was proposed by the author in the previous study. It establishes the general relation between microalgae growth rate and influence factors, which is written as [27]

$$Mg_x = \frac{0.664}{5} \frac{(N_s - N_\infty)}{N_\infty} Re_x^{1/2} \, Re_L^{1/2} \, Sc_N^{2/3} \tag{22}$$

for a laminar parallel flow, where $N$ represents the microalgae concentration. The physical meaning of the Microalgae growth number can be interpreted as the ratio of microalgae growth diffusivity and cell diffusivity. For similar microalgae growth process, the similarity principle requires the characteristic number, i.e., the Microalgae growth number to be as follows:

$$Mg' = Mg'' \tag{23}$$

In the above, the similarity of microalgae growth has been fully analyzed. Nevertheless, it is equally important that the similarity of the optical constants of microalgae are considered from an electromagnetic perspective. The temporal scaling function of absorption and scattering can be expressed as in [28].

$$Z_a(t_a, t_s) = \frac{k(t_a)\overline{d_{t_a}^3}}{k(t_s)\overline{d_{t_s}^3}} \tag{24}$$

$$Z_s(t_a, t_s) = \frac{n(t_a)(n(t_a)-1)^2 \overline{d_{t_a}^3}}{n(t_s)(n(t_s)-1)^2 \overline{d_{t_s}^3}} \tag{25}$$

where the $n$, and $k$ denote the refractive and absorption index, respectively, which are generally functions of wavelength, $t_a$ and $t_s$ represent the arbitrary growth time and time to reach the stationary phase, respectively. The bar represents the average of the integral over the cell diameter to the power of 3 with a distribution. Using the electromagnetic scattering theory, it has been proven that the temporal scaling function of time-dependent absorption and scattering cross-sections is largely independent of wavelength [28]. Therefore, the conclusion of temporal scaling function equality for similarity microalgae growth can be easily extended to the Vis-NIR spectral region. Furthermore, we can obtain the following relations for the time-dependent optical constants of microalgae

$$\left[\frac{k(t_a)\overline{d_{t_a}^3}}{k(t_s)\overline{d_{t_s}^3}}\right]' = \left[\frac{k(t_a)\overline{d_{t_a}^3}}{k(t_s)\overline{d_{t_s}^3}}\right]'' \tag{26}$$

$$\left[\frac{n(t_a)(n(t_a)-1)^2 \overline{d_{t_a}^3}}{n(t_s)(n(t_s)-1)^2 \overline{d_{t_s}^3}}\right]' = \left[\frac{n(t_a)(n(t_a)-1)^2 \overline{d_{t_a}^3}}{n(t_s)(n(t_s)-1)^2 \overline{d_{t_s}^3}}\right]'' \tag{27}$$

Assuming the cell size distribution is unchanged during the microalgae cultivation [29], the relations can be rewritten as follows:

$$\left[\frac{k(t_a)}{k(t_s)}\right]' = \left[\frac{k(t_a)}{k(t_s)}\right]'' \tag{28}$$

$$\left[\frac{n(t_a)(n(t_a)-1)^2}{n(t_s)(n(t_s)-1)^2}\right]' = \left[\frac{n(t_a)(n(t_a)-1)^2}{n(t_s)(n(t_s)-1)^2}\right]'' \tag{29}$$

Here, the equations mean that the time-dependent absorption index is linearly proportional for similarity growth of microalgae. Moreover, for similarity growth, the refractive index is linearly proportional by the expression $n(t_a)(n(t_a)-1)^2$. These relations are important to predict the optical constants varying with growth time for similarity growth of microalgae.

## 3. Conclusions

The objective of this letter is to study the similarity of microalgae growth. Besides the heat and mass transfer, the growth phenomenon is a crucial characteristic of microalgae cultivation. In accordance with the principle of similarity at the foundation of the similarity of heat and mass transfer, the characteristic number of microalgae growth and the temporal scaling function of extinction, absorption, and scattering should be equal. Moreover, the similarity relations of the time-dependent optical constants of microalgae are derived, which is crucial to understand the growth of microalgae. The group of characteristic numbers in this letter give the condition for similarity growth of microalgae. It is useful for designing and guiding experiments on microalgae cultivation.

**Declarations**

**Ethics approval and consent to participate**

No human or animal rights are applicable to this study.

**Availability of data and material**

The datasets analyzed during the study are available from the corresponding author upon reasonable request.

All data generated or analyzed during this study are included in this published article.

**Competing interests**

The author declares no competing interests.

**Funding**

This work was supported by Jiangxi Provincial Natural Science Foundation (No. 20212BAB214060) and Nanchang University, which are gratefully acknowledged.

**Authors' contributions**

The work presented here was completely carried out by Chunyang Ma. It includes the conception and design of the study, interpretation of the results, presentation of the article, and critical revision of the manuscript as well as final approval of the manuscript.

**Acknowledgements**